\documentclass[12pt]{article}
\usepackage{graphicx}
\usepackage{amssymb,amsmath,amsfonts,palatino,amsthm}
\usepackage{amssymb}
\usepackage{epstopdf}
\newcommand{\f}{\begin{equation}}
\newcommand{\ff}{\end{equation}}

\begin{document}

\title{Temporal naturalism }
\author{Lee Smolin\thanks{lsmolin@perimeterinstitute.ca} 
\\
\\
Perimeter Institute for Theoretical Physics,\\
31 Caroline Street North, Waterloo, Ontario N2J 2Y5, Canada}
\date{\today}
\maketitle

\begin{abstract}

Two people may claim both to be naturalists, but have divergent conceptions of basic elements of the natural world which lead them to mean different things when they talk about laws of nature, or states, or the role of mathematics in physics.  These disagreements do not much affect the ordinary practice of science which is about small subsystems of the universe, described or explained against a background, idealized to be fixed.  But these issues become crucial when we consider including the whole universe within our system, for then there is no fixed background to reference observables to.   I argue here that the key issue responsible for divergent versions of naturalism  and divergent approaches to cosmology is the conception of time. One version, which I call temporal naturalism, holds that time, in the sense of the succession of present moments, is real, and that laws of nature evolve in that time.  This is contrasted with timeless naturalism, which holds that laws are immutable and the present moment and its passage are illusions.  I argue that temporal naturalism is empirically more adequate than the alternatives, because it offers testable explanations for puzzles its rivals  cannot address, and is likely a better basis for solving major puzzles that presently face cosmology and physics.

This essay also addresses the problem of qualia and experience within naturalism and argues that only temporal naturalism can make a place for qualia as  intrinsic qualities of matter.

\end{abstract}

\newpage

\tableofcontents


\section{Introduction}

The subject  of this essay is {\it naturalism,} the philosophical stance most closely associated with what might loosely be called the scientific world view.  My particular aim is to point out the role that our conception of time plays in the scientific conception of nature and to advocate a particular form of naturalism based on an embrace of the reality of time, in the sense (to be defined carefully below) of the present moment and its passage.   

I will begin by defining the form of naturalism I advocate,  {\it  temporal naturalism}, and distinguish it from its main rival, {\it timeless naturalism}, the view that what is really real is the whole history of the universe, taken as one.  Timeless naturalism includes the block universe interpretation of general relativity.  One  variant of it is the naturalism based on a timeless {\it pluralism of moments} advocated by Julian Barbour.  I then show that the choice of naturalisms has consequences for the practice of science, particularly cosmology, due to the implications of the nature of time for the conception of laws.

Timeless naturalism is similar to the view philosophers call ``eternalism"
 and temporal naturalism has elements in common with the philosophers' ``presentism", but my categories differ from the older ones, among other reasons, because of an emphasis on the nature of law with regard to time.   

This essay is part of a larger project whose aim is to radically reconfigure the practice of science on a cosmological scale to admit three theses: 1) the reality of time, 2) the evolution of laws with respect to that time and 3) the uniqueness of the single causally closed universe that unfolds in time.  This project was conceived with Roberto Mangabeira Unger and  its main vehicle is a book jointly written with him\cite{SU} to which my book, Time Reborn\cite{TR} may serve as an introduction.  
It is the combination of the three theses that makes this not just a rerun of the old presentist-eternalist debate.  In particular, a major  claim of
\cite{TR,SU} is that temporal naturalism has a much larger range of empirical adequacy than its rivals because only it allows a conception of laws which can evolve in time.  This, we argue, is necessary if we wish the choice  of laws to be explicable on the basis of hypotheses that are falsifiable by the results of doable experiments.  

The present essay summarizes the main arguments of \cite{TR,SU} and then advances the program in two directions.  

\begin{enumerate}

\item{} Recent work with Marina Cortes\cite{ECS1,ECS2} has emphasized three aspects of a reconfiguration of cosmology within temporal naturalism: i)  framing the fundamental laws as asymmetric and irreversible in time, ii) drawing out the consequences of an insistence on the uniqueness of elementary events (an under explored consequence of Leibniz's Principle of the Identity of the Indiscernible) and iii) finding a role for energy and momentum as necessarily intrinsic quantities within a relational universe.  One consequence\cite{ECS1,ECS2} is an elegant solution to the longstanding problem of getting classical space-time to emerge from the semiclassical limit of a theory of causal sets\cite{CS}.

\item{} As emphasized by Galen Strawson\cite{GS} and Thomas Nagel\cite{TN},
naturalism must have a role for qualia and experience, if it is to live up to its defining aspiration that everything that we know that exists is part of the natural world.  I argue below that only temporal naturalism  can accommodate qualia and experience as intrinsic qualities of events in nature.  

\end{enumerate}

\subsection{What is naturalism?}

Let me start with a definition:

\begin{quotation}
Naturalism is the view that all that exists is the natural world that is perceived with, but exists independently of, our senses or tools which extend them; naturalists also hold that science is the most reliable route to knowledge about nature. 
\end{quotation}

 This is a good first try at a definition but its simplicity hides ambiguities and traps.  Unless we are idealists we do not believe that all that exists are our perception.  What we believe is that our senses give us evidence for the existence of a natural world, which can be learned about through our sensations but which exists independently of them.   

However our senses, and the experiments and observations which we carry out to extend them, only give us direct acquaintance with the qualia which are the sensory elements of our experience.   They do not give us immediate acquaintance with, or direct knowledge of, the rest of the natural world.   They can then only provide evidence for hypotheses which we make concerning the natural world.  Thus, as naturalists we are constrained to deal in indirect knowledge of the object of our study and we must be always conscious that this knowledge is incomplete and never completely certain.  But since we believe all that exists is the natural world we must admit that incomplete and tentative knowledge is the best that can be had concerning what exists.

Because of this naturalists can hold quite strikingly different views about nature-and still be naturalists.  For example, many naturalists believe that everything that happens in nature is governed by universal and unchanging laws.  But one doesnÕt have to believe this to be a naturalist-because we must admit the possibility that experiment could provide evidence for phenomena that are governed by no definite law.  For example, if we believe that no hidden variables theory determines the precise outcomes of measurements on quantum systems for which quantum mechanics only gives probabilistic predictions, then we believe there are phenomena that are not law governed at all.  Indeed if we follow Conway and Kochen\cite{CK} then quantum phenomena are in a well defined sense free\cite{precedence}.  Or, if we believe the standard big bang cosmology expressed in the context of classical general relativity then we implicitly believe that no law picks the initial conditions of the universe.  Or to put it another way, no law governs which solution to the equations of general relativity is somehow uniquely blessed with describing the actual history of the universe.  

Another thing that some, but not all naturalists believe is that everything that exists in the natural world can be completely described by the language of physics.  There are varieties of positions held with respect to emergence and reduction; but it is quite reasonable to believe that matter is made out of elementary particles which obey general laws, but that complex systems made out of many atoms can have emergent properties not expressible in or derivable from the properties of elementary particles.

\subsection{Naturalism through the looking glass}

Many naturalists hold beliefs about the natural world that are more firmly held and expressed than the tentative nature of scientific hypotheses allows.  These are often beliefs of the form:

\begin{quotation}
{\it ÒOur sense impressions are illusions, and behind them is a natural world which is really XÓ.} 
\end{quotation} 

Such a view can either be an ordinary scientific hypothesis or a metaphysical delusion, depending on what X is asserted to be.  
When X is a statement like {\it ``made of atoms"}  this is an innocuous scientific hypothesis which carries little metaphysical baggage and is, in fact, very well confirmed by diverse kinds of experiments.  (But this was of course not always the case.)  
But statements of this form can be traps when X is a big metaphysical assertion which goes way beyond the actual evidence. 

A common and widely believed example is the claim that {\it X (the universe) is really is a timeless mathematical object\cite{Max}.}  Whether that mathematical object is a solution to an appropriate extension of general relativity or a vector in an infinite dimensional space of solutions to the Wheeler-deWitt equation of quantum cosmology there is a big stretch from a statement of the form,

\begin{quotation}
{\it ÒSome experimental evidence concerning a specified range of phenomena is well modeled by a mathematical object, O,Ó} 
\end{quotation}  
which is a statement which might or might not be supported by evidence and a metaphysical assertion that {\it ÒThe universe is really a mathematical objectÓ},  which is not by any reach of the imagination an hypothesis that could be tested and confirmed or falsified.

What is troubling is that statements of the form, {\it ÒExperience is an illusion, the universe is really XÓ}  are common in religion.  When naturalists make statements of this kind, they are falling for what might be called the transcendental folly. They are replacing the concrete natural world by an invented conception, which they take to be ``more real" than nature itself.  Thinking like this turns naturalism into its opposite.

Much that passes for naturalism and physicalism these days are instances of transcendental folly.

A symptom of the transcendental folly is the move from {\it ``Sense impressions give unreliable knowledge of nature, nature is instead truly X"} to 
{\it ``Sense impressions are incompatible with the concept that the world is X"}, so qualia must not exist.  But the one thing we can be sure of is that qualia exist.   Therefor, as Galen Strawson\cite{GS} and other philosophers of mind emphasize, if we are naturalists and believe everything that exists is part of the natural world then qualia must be also part of the natural world.   The right statement-if we are naturalists-must be:

\begin{quotation}
 {\it ``X may provide a good description of some class of observations of the world, but the world cannot be X exactly because qualia are undeniably part of the world and X are not qualia."}
 
 \end{quotation}

This is especially the case when X is "a mathematical object."  

\subsection{Naturalism is an ethical stance}

Part of my definition of naturalism refers to science as the most reliable path to knowledge about nature.  Note that my definition doesn't claim that science is the only path to knowledge, nor does it call on or require that there be a scientific method.  So I have to flesh out the definition by explaining what I mean by science as a route to knowledge about nature.   Most importantly, I need to emphasize that while, as Feyerabend convincingly argued, there is no scientific method, science is most fundamentally defined as a collection of ethical communities, each organized around a particular subject.  An ethical community is a community, membership in which is defined by holding and following of certain ethical principles.  In \cite{TTWP,TR} I argued that the scientific community is defined by two ethical principles.  To quote from \cite{TR}

\begin{quotation}

{\it ``Scientific communities, and the larger democratic societies from which they evolved, progress because their work is governed by two basic principles.

\begin{enumerate}

\item{} When rational argument from public evidence suffices to decide a question, it must be considered to be so decided.

\item{} When rational argument from public evidence does not suffice to decide a question, the community must encourage a diverse range of viewpoints and hypotheses consistent with a good- faith attempt to develop convincing public evidence.

\end{enumerate}

I call these the principles of the open future. "}

\end{quotation}

Naturalism is then in part an ethical commitment.  



\subsection{Relationalism}

I am a Leibnizian, which I take to mean that I find the following of his principles to be very helpful to frame the search for a correct cosmological theory. 

\begin{itemize}

\item{} Principle of (aspiration for) sufficient reason. 

\item{} Principle of the identity of the indiscernible. 

\item{} Principle of causal closure: the universe contains all its 
causes. 

\item{} Principle of reciprocity:  if an element of nature, A, can influence change in an element B, the reverse must also be the case.

\end{itemize}

Leibniz's principles underlie the philosophy of {\it relationalism}, which claims that at least some physical quantities, particularly space and time, are defined only by the network of relationships that weave physical reality into  a whole.
However I am not a relational purist, i.e. someone who believes all physical quantities are relationally defined.  There can be intrinsic qualities as well and among these, I will argue below, are energy and momentum.  Another class of intrinsic qualities I will discuss are qualia.  

I call the principle of sufficient reason an aspiration because I believe its use for science lies in motivating us to attempt always to invent hypotheses and theories that will increase our understanding of the reasons the universe is the way it is.  General relativity is the best example of its power, and it continues to underlie developments in quantum gravity as the desire that our theories be ``background independent" lies in our understanding that there can be no sufficient reason for background structure in a fundamental theory.  And, as I explain here, the desire for sufficient reason underlies our insistence that there must be a causal explanation for the universe's choice of laws and initial conditions.

At the same time, the fact that it is generally good advice to seek to come closer to sufficient reason in no way commits us to the belief that sufficient reason must be completely obtained by a cosmological theory to judge it a success.  In fact, there is a tension between the demand for total sufficient reason and the insistence that there is a single universe, unique in all its particularity.  We must admit the possibility that there may be many details of how the universe is that are simply the way it is.  Indeed, if the current models of inflationary cosmology are right then the details of the structure in the universe are seeded by quantum fluctuations of fields during inflation which, on the standard interpretations of quantum theory, are truly random and hence have no sufficient reason.   Nor is it obvious it would increase sufficient reason to explain these in terms  of hidden variables as this would likely just add to the burden carried by the initial conditions.  And there is a further tension between sufficient reason and the demand that the future be partly open, which is also part of what tempural naturalism endorses.  

It is best then to decouple the heuristic usefulness of the search for sufficient reason from the knotty question of whether there is ultimately complete sufficient reason by speaking of a {\it Principle of differential sufficient reason: it is always good advice to seek to increase our understanding in ways that bring us closer to sufficient reason.}

The principle of reciprocity was introduced by Einstein as part of the motivation for general relativity.  I will widen its application to criticize the idea that laws of nature are immutable if they act to cause events to happen but in turn cannot be changed or modified by anything that happens inside the universe.  However, note that I will allow an important exception to this principle as I will posit that the past influences the present and future, but the reverse is not true; no event  in the past can be affected by anything that happens in its future.   

Finally,  I will apply the Principle of the identity of the indiscernible not, as is sometimes done, in the context of possible worlds-where it is false, but as a criteria our single unique existing universe must satisfy.  Thus, while it is possible to imagine possible universes that have symmetries that render two or more distinct events indiscernible, I will argue that such symmetries must be absent in the one real universe.

\section{Naturalism,  time and laws}

In this essay I am interested to tease out distinct kinds of naturalists which differ because of different views as to the nature of time as well as the relationship between time and laws.  I will mainly discuss two views which I will call timeless naturalism and temporal naturalism.  

I discuss each in turn.

\subsection{Timeless naturalism}

\begin{itemize}

\item{} { \bf Timeless naturalism} holds that the experience of moments of time and their passage or flow are illusions.  What really exists is the entire history of the universe taken as a timeless whole.  {\it ``Now"} is as subjective as {\it ``here"}  and both are descriptions of the perspective of an individual observer.  There are, similarly, no objective facts of the matter corresponding to distinctions between past, present and future.   

Timeless naturalism is closely related to the block universe interpretation of general relativity.  That interpretation  is supported by arguments such as that of Putnam\cite{Putnam} who hold that the relativity of simultaneity implies that any property of a moment must be shared equally by all moments. 

Timeless naturalism holds that the fundamental laws of nature are timeless and immutable.   

To avoid confusion we should recognize that there are several different versions of the block universe idea in the physics and philosophy literature.  The {\it physicists' block universe} assumes that the universe's history is the result of deterministic and immutable laws acting on initial conditions, and this is what I will refer to when I speak of the block universe.  Within this conception of the universe nothing can ever be truly novel because any observable measuring a property of the universe at a future time can be expressed as a function of observables at any earlier time, by inverting the equations of motion.  The {\it special relativistic block universe} is a special case of the physicists block universe where simultaneity is relative so the argument of Putnam against presentism can be run.   Even stronger is the {\it general relativistic block universe}, where the assumed spatial compactness plus space-time diffeomorphism invariance implies that any observable at one time is {\it gauge equivalent} to observables at any other time, i.e. observables at different times are mathematically different but physically equivalent ways to describe the same facts.

Some philosophers defend a very much weaker notion of a block universe, which is just the collection of all facts that will ever ``be the case" at sometime or another, irrespective of whether the laws are deterministic or not, or changeable or not.  Or, as it is sometimes put,  ``wait to the end time and then represent all that has been the case" in a single mathematical representation of the universe's entire, now past, history.   This is a metaphysically fanciful notion whose relevance to any issue of physics can be denied by just saying no to the existence, interest or validity of the viewpoint of an observer, not only outside the universe, but also beyond it temporally and causally.   But allowing this {\it philosophers' block universe} into discussion, it is clear that, as is claimed by philosophers, there is no contradiction between it and any notion that laws evolve.  This is true, but it is not the notion of a block universe that physicists speak. So when I use the words ``block universe" in this paper I mean the physicists' block universe and, unless otherwise specified, I mean it in the strong form of the general relativistic block universe.

The physicists' block universe  is then a description of the timeless reality which comprises the history of the universe, taken as one, that timeless naturalism asserts is what is real.   

\subsubsection{Can timeless naturalism recover our experience of the world?}

To be at all a candidate for a description of our world, timeless naturalism has to answer a simple question: how does the sense of experiencing the world as a set of moments arise from a world with no conception of time?  Carnap addressed this while recounting his conversations with Einstein on the great scientist's disquiet with the absence of a conception of the now in physics,

\begin{quotation}

{\it   ``. . .  the peculiarities of manÕs experiences with respect to time, including his different attitude towards past, present, and future, can be described and (in principle) explained in psychology\cite{Carnap}."}
\end{quotation}

I believe a few moments thought suffices to convince us that it will require more than a call on psychology to satisfactorily answer this question.  Two very different kinds of claims would need to be made and defended before Carnap's aspiration can be fulfilled.  

First,  it has to be established that there are any facts in the block universe corresponding to events that are local in both space and time-as is the case with our experiences.  For once it is asserted that what is real is the whole history of the universe, taken as one, it is not obvious that there are any purely local facts or observables.  The second point that must be argued is that the block universe picture can explain all the features of our experiences.

Some proponents of {\it eternalism} (a view related to, but not identical to what I am calling timeless naturalism\footnote{Because eternalize makes usually no claim as to the status of laws, whereas I am taking a belief that the universe is governed-at a fundamental scale-by a timeless  and immutable law as part of the definition of timeless naturalism.}) hold that the timeless universe contains ``temporal parts" which are local in time, and  that some of these  correspond to the moments we experience.  Others hold that the fact that we experience the world through moments and remember the past but not the future is an evolutionary adaptation which arises to make {\it information gathering and utilizing subsystems} more efficient and hence more fit\cite{Hartle}-although the extent to which this argument is not circular because it assumes an evolution {\it in time} is a subtle question.  

But it can't be this simple because whether or not one can simply assert that there are local observables in the block universe picture depends on the physics.  Newtonian physics is local in time, and so are its observables, but things are not nearly so simple in general relativity.  While it is true that Einstein's field equations are local in space-time, the observables are generally not, because reality corresponds not to a solution of the field  equations but to an equivalence class consisting of the infinite number of solutions rendered equivalent under diffeomorphisms.  

This raises an important point I will develop below, which is that to refer to any local event in general relativity requires giving a complete specification of information sufficient to distinguish it from all the other events in the space-time.  This is required to refer to an event uniquely in a diffeomorphism invariant fashion-and it can also be seen as a realization of the principle of the identity of the indiscernible.   The key point to note is that any such local observable must then be defined relationally and it is also contingent because the conditions that uniquely specify a particular event may not obtain in every solution of the field equations.  I raise below the question of whether our subjective experience can ever be associated with such a relational and contingent description.

If this worry  can be resolved, one would then have to face the second goal which is to explain how all the features of our experience can be reproduced within the block universe.   There are recently sophisticated attempts to do this using recent results of cognitive science\cite{Jennan-cogsci,LaurieP-cogsci}.  These are impressively detailed but even if we grant they succeed, their relevance rests on the adequacy of the answer to the first point, which I will argue cannot be satisfactorily answered.

A related issue is causality.   The temporal parts of a timeless universe are normally taken to be ordered by causal processes.  But  Julian Barbour points out that when quantum mechanics is applied to cosmology by means of the Wheeler-deWitt equations, causality and order disappear and we are left only with the set of possible moments to which the wave function of the universe assigns probability amplitudes.   This gives rise to the version of timeless naturalism we may call Barbour's {\it instantaneous pluralism of moments\cite{TEOT}.} This is an interpretation of the timeless Wheeler-deWitt equations of quantum  cosmology,  according to which what exists is a vast collection of moments, which exist all together timelessly.  On this view there is no order to the moments, nor is there any structure which distinguishes one moment 
as the present moment.  

According to Barbour\cite{TEOT} there is a single fundamental law, the Wheeler deWitt equation, which determines the distribution of different moments in the collection.

\subsection{Temporal naturalism}

\item{}{ \bf Temporal naturalism}, on the other hand, holds that all that is real is so at a moment of time, which is one of a succession of moments.  Because of the qualification that the real moment is one of a succession, this view is not equivalent to what the philosophers call presentism.  In particular,  there are two kinds of things that are real at each moment.  The first class consists of events which are  real at that moment.  The second kind of real thing is the process which continually brings forth new events from present events.   This can be called the {\it activity of time\cite{ECS1}.}  This process underlies causation which we take to be prior to laws.  

According to this view, the future is not real and there are no facts of the matter about it.  The past consists of events or moments which have been real.  Even though past events are no longer real there can be facts of the matter about them.  Some of these can be decided based on present evidence of  past moments,  such as fossils, structures, records etc.  

The status of the past is a tricky issue for temporal naturalists because one is simultaneously asserting that the past is not real and yet statements about the past can be true or false.  One can certainly  evaluate the likelihood for statements about the  past to be true based on present evidence.  But I want to believe that these statements are true or false independent of whatever present evidence there may be for them.  It is the case that the dinosaurs roamed the earth and will remain the case even after the Earth disappears in the Sun's expansion to a red giant.  But does this mean that I must believe that statements can be true or false even if they refer to nothing that exists?  The answer is yes, this  is an aspect of temporal naturalism.  This is no problem and just serves to underlie the extent to which the concept of existence must be modified if one takes the view that all that exists is the present.

Temporal naturalism further holds that the laws of nature can and do evolve in time and that, while there may be principles which guide their evolution, the future may not be completely determined.  This is consistent with the claim that there are no facts of the matter about the future.

\end{itemize}

\subsection{Distinguishing the two forms of naturalism}

The differences between these different versions of naturalism are brought out if we ask 
several questions: 

\begin{enumerate}

\item{}{\it How does the organization of our experience in terms of a one dimensional succession of moments arise?}

This is a surprisingly hard question for a timeless naturalist, because if all that is real is the whole history of the world taken as one, then that history should be no more broken into moments then a sculpture by Picasso should consist of points that are intrinsically ordered.  As already mentioned, locality is neither obvious nor simple in the general relativists's block universe.   Moreover the same shape, cast in bronze, admits an infinity of orderings-what picks out which is fundamental?  Indeed even given that the block universe has local properties, who tells us it is those which are the correlates of experience?  A general relativistic universe also has observables non-local in space and time-why are these not the correlates of experience\footnote{Indeed is a panpsychist constrained to assert experience is correlated with all physical observables of the correct theory, whether these are local or non-local?}?  Indeed,  why the world should be experienced as a set of anythings rather than simply all at once-as it is claimed to exist all at once-is a very hard question for a timeless naturalists-one that to my knowledge has never been satisfactorily answered.  We can follow Chalmers and call this the {\it hard question of time.}  


But that is not all.  Once the hard question of time is addressed (or conceded),  a timeless naturalist has to explain what is special about timeline trajectories: Why aren't space like trajectories the ones which are experienced as flows of moments?  (Some timeless naturalists assert they could be.)   Similarly, what is special about  a one dimensional order?  Why don't our experiences of the world organize themselves over two dimensions-as if we were all strings?

Temporal naturalism solves the hard problem of time by definition because it explains that we experience the world as a succession of moments because the world really exists as a succession of moments.  The hard problem that temporal naturalism has to address is why anything persists, i.e. why there appear to be causal connections between past and present moments.  Causality has to be fundamental and-as we shall see-prior to general laws.  


But even is temporal naturalism solves the hard problem of time by definition, this cannot be the end of the story.  The temporal naturalist has also to explain all the features of  time as perceived by us and some of these, like the thickness and plasticity of the experienced moment, as well as the experience of flow or passage, may involve hard cognitive science.  Indeed, it seems unlikely that the moment can be thin, so that the moments the world passes through must be thick and structured\cite{ECS1}.  

\item{}{\it Why do we remember the past but not the future?}

This is a hard problem for a timeless naturalist, who has to begin by conceding that there is no fundamental distinction between the future and the past and so admit that the different ways in which our universe appears to be highly asymmetric in time are contingent consequences of our universe being in a very special state\footnote{While most fundamental theiries under consideration are symmetric under time reversal, there are rare exceptions; among them are the spontaneous collapse extensions of quantum mechanics which posit time asymmetric stochastic laws for wave function collapse\cite{GRW}. David Albert argues that these could provide an explanation for the arrows of time that does not rely on the past hypothesis\cite{Albert}.  But given that it requires a preferred global time and rejects unitary, deterministic dynamics in favour of stochastic evolution of the wave function, the GRW theory can be seen to fit within temporal naturalism.}.  Were the universe in a different state we might ``remember" -that is have fossils of-both the future and the past.  In other universe there would be memories or fossils neither of the past nor the future because they would be in thermal equilibrium in which improbable non-random structures are highly improbable.   Moreover, there is a well defined sense in which in both general relativity and Newtonian cosmologies universes in equilibrium are vastly more probable-given any reasonable time symmetric measure on the space of solutions-than are universes out of equilibrium.  This leads to the  nearly absurd claim that, on our best understanding, the single universe that exists is highly improbable.  

For a temporal naturalists this question has a simple answer: because the past was real and caused the present, whereas the future is not presently real.  Moreover since a temporal naturalist conceives the history of the world as fundamentally asymmetric in time, there is no obstacle to hypothesizing that the fundamental laws are time asymmetric.  This opens up the possibility of explaining the strong time asymmetry of our universe with time asymmetric fundamental laws\cite{ECS1}\footnote{As indeed Penrose proposed when he posited his hypothesis that the Weyl curvature tensor vanishes at initial but not at final singularities\cite{Penrose}.}.

\item{} { \it Can we speak preferentially about the present moment? Can we use the word ÒnowÓ to 
refer to a fact about the world? }

In temporal naturalism we can because the present moment is real and there can be facts about it.  In this version of naturalism it suffices to say, ``now". 
This refers simply and non-contingently to the moment which is now present,  which we might call the presently present moment. 

In timeless naturalism ÒnowÓ is the same as ÒhereÓ; it is always a stand in for a relational statement which refers to an observation in a way that indexes it to timeless observables. ie:  ``by now I mean at the same spacetime event as a particular physicist named Lee Smolin, sitting at his dining room table at a time when the clock reads 6 am and the calendar reads August 12, 2013."  We can call this relational  addressing. It is the replacement of a use of ÒnowÓ with a contingent and relational statement that refers only indirectly to a moment by a relational feature that distinguishes it from all other moments in the history of the universe. In
other words, this is the detailing of a local statement in  time by a complicated  observable which is non-local in time, in the sense   
that its concrete realization would require a function of variables all  over spacetime. 
 
As I will discuss below, such a reference to a moment is necessarily  contingent, i.e. I may have been still in bed at the stated time. 

The same is the case in Barbour's moment pluralism, references to a particular moment must always be made relationally and contingently, by describing what someone ``present" in that moment would be experiencing.  One difference is that the ``heap of moments" is envisioned to include all possible configurations of the world so that there are moments where I am in bed when the clock reads 6 am and the calendar reads August 12, 2013, moments when I am at a desk (in all the possible houses where my life might have taken me.)  Of course in most moments I don't exist at all.  

Thus, in a timeless natural world there is no use of the presently present moment. Instead any 
statement abut ``now", apparently local in time, is to be replaced by a non-local observable that relationally and contingently addresses that moment. 

In temporal naturalism, ÒnowÓ is an intrinsic property, in the sense, to be developed below, that it is not relational. It requires no further specification. 

In timeless naturalism and moment pluralism, ÒnowÓ is not intrinsic, it is relational and can be defined only 
indirectly through the specification of a very complex, contingent and highly non-local observable. ``Now" has no meaning. ``Now in the SW corner of Trinity Bellwoods park at noon on the first Sunday in September, 2115" does. 
 
\item{} { \it Can we give truth values to facts apart from those that refer to now? }

In temporal naturalism, we can give truth values or probabilities to hypotheses about the past.  The truth values are objective but to evaluate them requires  evidence available in the present moment. No 
statement about the future has a present truth value. 

In timeless naturalism statements with truth values do not distinguish past, present or 
future. For example, in a block universe interpretation of general relativity, all physical observables are non-local in time.  The same is true in moment pluralism.  

\end{enumerate}

\section{Time and laws}

The choice between temporal and timeless naturalism has implications that span many of the issues facing modern science and go deep to affect how we conceive of the naturalist project.  The most immediate implication is to our concept of a law of nature.

To a timeless naturalist no fundamental law can give a privileged status to the present or refer to the distinction between past, present and future.  As there is no objective distinction between past and future any law that is truly fundamental must be timeless and immutable.

\subsection{The Newtonian paradigm}

The standard for explanation in physical science is the Newtonian paradigm. It is the formal structure shared by classical mechanics, quantum mechanics, general relativity, quantum field theory, quantum gravity and some models of computation.  
The key to the Newtonian paradigm is the separation of explanation into two parts:  Laws and initial conditions. 

To set up a theory in the Newtonian paradigm one must first specify a subsystem of the universe which can be plausibly be treated, to a reasonable approximation, as an  isolated system, and then answer two questions about it: 
 
 \begin{enumerate}

\item{} {\it What are the possible states of  the system at a fixed time?  These states, it is assumed, can be prepared by an experimenter at an initial time or determined by the outcome of a complete measurement at a later time\footnote{For reasons that will shortly become clear we prefer this operational notion of a state to the weaker notion of ``all that is the case about  a system."}.}

\item{} {\it How do these states change in time?}
 
 \end{enumerate}

These two questions correspond, respectively, to what is called the kinematics or dynamics of the system.  

The answers to these questions must be timeless, ie not themselves change in time.
These answers are coded in the following formal, mathematical structures.

\begin{itemize}

\item{}A space of states, $\cal P$ whose points represent the possible states of the system, that is the results of a preparation or complete measurement.  In classical mechanics this is the phase space (pairs of configuration and momentum variables), in quantum mechanics this is the Hilbert space.

\item{}A vector field, $h^a$ on $\cal P$, or its integral, which is a foliation of $\cal P$ by a family of paths, $\gamma (t)$, exactly one of which passes through each point $p \in {\cal P}$.

\end{itemize}

The Newtonian paradigm joins to this formal structure an experimental practice within which the object that is studied and formally represented is a subsystem of the universe, idealized as an isolated system.  The observers remain outside the system along with clocks\footnote{There are very special classical systems in which the time parameter can be parameterized by internal clocks, but their use does not strongly affect the present arguments.} used to measure duration and measuring instruments used to determine and, in some cases, prepare the state of the system.  When properly used, this combination of experimental practice and formal representation underlies the dramatic success of physics since Galileo, Kepler and Newton.  

\subsection{The proper use of the Newtonian paradigm}

The proper use of the Newtonian paradigm, by which I mean the use that matches up with experimental practice, is to describe records of past observations of isolated 
systems.  One assumes the existence of a system, {\bf S}, isolated from the rest of nature except for its interaction with a set of measuring instruments, controlled by an observer outside the system.  There also must be a clock outside the system.

States then correspond to possible preparations or complete measurements, as done by an observer using instruments external to the system. Evolution is defined with respect to a clock external to the system.  To proceed one makes  a series of measurements at times $t_1, t_2, \dots $, to determine the state at 
each time.  This resulted in a record, $ \{p (t_1), p(t_2), \ldots \}$. 

This record, once made, is static, it doesn't change in time.  It is therefor entirely appropriate to represent the record by a curve $\gamma (t)$, which coincides with the entries in the record at the stated times.  This is a mathematical object,  which is also unchanging in time. 

\subsection{The cosmological fallacy}

It is however fallacious to infer from this that nature is a mathematical object or is ``really timeless."   The reason is that the context of referring to an isolated system controlled or selected by an external observer is essential. 
This dependence on the context of a subsystem of the universe blocks any broad metaphysical conclusions being drawn from the success of the Newtonian paradigm, because any such deduction would require the application of the paradigm to the whole universe.

To see why the success of the paradigm is due to its domain of validity being restricted to subsystems, note that there are several aspects to the matching between experimental practice and the formal structure of the Newtonian paradigm.

\begin{itemize}

\item{} There are many possible trajectories because there are many possible initial conditions that may be chosen by the experimentalist. Thus, two systems, otherwse identical in terms of their constituents, the laws they obey and the boundary conditions and outside influences they are subject to, can have different histories.  Furthermore, two histories cannot share any states, because there is a unique history through every point $p$ of the state space, $\cal P$.  These are far from obvious features of natural systems and show the power of the Newtonian paradigm.

\item{} Because the isolated system is a small subsystem of the universe, it can be repeated many  times with different initial conditions. The experimenter can use their freedom to do the experiment many times to vary the initial conditions and, by doing so, test hypotheses as to the laws.
 
\item{} So the absolute separation of laws and initial conditions, and thus of laws and states, is tied to the empirical context of studying small subsystems of the universe. 

\item{} The time parameters $t$ in the records and trajectory corresponds to the reading of a clock outside the isolated system.

\end{itemize}

However-and this is a key point- the good fit with an empirical context falls apart if we try to apply this paradigm to the universe as a whole. 

\begin{itemize}

\item{} Unlike the usual case there is only one system to study and it happens once.

\item{}There is only one unique history which is the history of the actual universe. How is that determined? It cannot be selected or imposed by any observer.  But a choice of initial conditions for the unique universe must nonetheless somehow be made.  How?

\item{}  There is no role for the other trajectories which are not chosen.  
 
Hence, when one attempts to apply the Newtonian paradigm to the universe as a whole, there is a mismatch.  It explains at once too little and too much. It explains too little because it cannot tell us why the one history that is realized in nature is distinguished thusly from the infinity of other trajectories which play no role in nature.  It explains to much because it explains an infinitude of potential, but unrealized, facts about the infinite number of  unrealized histories.  

\item{} There is no possibility of choosing or varying the initial conditions, hence of operationally 
separating the role of laws from that of initial conditions.   
 
This makes the application of theories to explain cosmological data different in important methodological ways than the application of theory either to data in the laboratory or in astronomy where one has many similar systems to study.   Rather than using the freedom to vary the initial conditions, or gather lots of data on similar systems with diverse initial conditions, to test hypotheses as to the laws, cosmologists are required to {\it simultaneously} test hypotheses about the laws {\it and} hypotheses about the initial conditions.  This leads in some real cases to a problem of underdetermination because it is impossible to separate the role of laws from that
of initial conditions in attempts to interpret cosmological data\footnote{Some examples of this problem are described in \cite{TR}.}.
 
\end{itemize}

To ignore this and attempt to scale up the Newtonian paradigm to the universe as a whole is to commit the cosmological fallacy. 
An important example of the cosmological fallacy and a false deduction from it is the claim that the right interpretation of the equations of classical or quantum cosmology is that the universe is timeless.  
Classical and quantum models of cosmology appear timeless because they arise from applying to a system with no external clock a method and formalisms whose empirical context requires an external clock. To deduce from the formalism of  classical or quantum general relativity that the universe is timeless is fallacious. 

\subsection{The cosmological dilemma}

A dilemma arises from the cosmological fallacy because the Newtonian paradigm, when properly applied to an isolated system in nature, must be seen as an approximation rather than an exact description.  This arises because when we regard a subsystem of the universe as an isolated system, we must ignore  interactions, which certainly are there, between degrees of freedom in the isolated system and degrees of freedom outside of its boundaries.  But no subsystem of the universe can be completely isolated-at worst, it is physically impossible to shield a system from the influence of gravitational waves coming from the outside\cite{noisolated}.  

One is tempted to remedy this problem by expanding the boundaries of the subsystem we regard as isolated by including more and more in it.  But if we try to take this process of enlargement to its necessary limit, and include the whole universe in the system, the explanatory context falls apart due to the inability to cleanly separate the roles of laws and initial conditions.   As  I have just discussed, we commit the cosmological fallacy.  

This is the {\it cosmological dilemma}\cite{TR,SU}.  The only solution to it is to {\it regard any application of the Newtonian paradigm as approximate.}  Hence, the mathematical precision which the Newtonian paradigm allows must necessarily coexist with an unavoidable limit to the precision with which the predictions of the theory may be compared with experiment.   Happily this coincides with the fact that theoretical physicists have learned it is best to regard our most successful theories, including QED, the standard model of elementary particles, and general relativity,  as {\it effective field theories.}  These describe a truncation of nature gotten by limiting the system to being modelled to a subset of the degrees of freedom of the universe.

The question on the table is then how can we extend physics to a description of the whole universe.  The proposal of this essay and the books it summarizes is that this can only be done by embracing the tenets of temporal naturalism.  As argued in the books, and as I will expand on further below, it is the mostly unconscious reliance on the metaphysical fantasies behind timeless naturalism, that is responsible for the current crises in cosmological theorizing.  The answer then to the end of science threatened by adoption of the multiverse is to embrace the reality of time and the evolution of laws.  

\section{The failure of timeless naturalism in the face of cosmological questions}
 
Timeless naturalism also fails because it renders key questions unanswerable. There are three key questions, each crucial for understanding the world on a cosmological time scale, which no theory couched in the Newtonian paradigm can satisfactorily answer.

\begin{enumerate}

\item{} Why these laws? 

\item{} Why these initial conditions?

\item{} Why so the universe so long out of equilibrium? 

\end{enumerate}

The Newtonian paradigm takes the choice of laws and initial conditions as input.  Furthermore there is no evidence that mathematics serves to limit the choices of possible theories sufficiently to explain most features of the known laws.  This is the case even if one restricts ourselves to unifications of the known forces within quantum theory\cite{LOTC}.  Hence it is impossible that the Newtonian paradigm could explain the choice of laws.  

Nor can the Newtonian paradigm offer any answer to the problem of the choice of initial conditions.  It is not enough to say that the big bang is represented in classical general relativity by a solution to general relativity with an initial cosmological singularity because there are an infinite number of such solutions and those that correctly describe our universe are quite atypical in the absence of inhomogeneities and black or white holes in the early universe.  Penrose characterized this by the hypothesis that the Weyl curvature tensor vanishes initially\cite{Penrose}.  Nor does inflation help because very stringent conditions on inhomoegeneities must be imposed for a solution to in late significantly.

Question 3 is a particular aspect of the initial conditions problem.  The question is why our universe is still out of equilibrium, still characterized by strong arrows of time, more than 13 billion years after its origin.  As Penrose has argued, part of the answer, within the Newtonian paradigm, must be that the 
entropy was initially so low that the second law has not yet increased it to its maximum at equilibrium.  This is necessary because the fundamental laws are believed to be invariant under reversal of the direction of time, so there can be no fundamental distinction between the past and the future.  

In \cite{LOTC}, I  called the why these laws problem, the {\it landscape problem.}  It was been much debated.  The current crisis in theoretical cosmology can largely be characterized as the failure of the method of explanation based on the Newtonian paradigm to answer these three key cosmological questions.

The different versions of naturalism offer very distinct responses  to this crisis.   These follow from their different roles each gives to the Newtonian paradigm and consequently to the concept of time on a cosmological scale.

\subsection{Timeless naturalism and law}

Timeless naturalism makes the following claims as to the nature of laws of nature.

\begin{itemize}

\item{}  The Newtonian paradigm extends to the whole universe.

\item{}  The distinction between states and laws is absolute. 

\item{}  Laws are timeless and immutable.

\item{}  The laws of nature are reversible in time because there is no fundamental distinction between past and future. 

\item{}  The history of the universe is isomorphic to a mathematical object. 

\end{itemize}

This is the usual picture most theoretical physicists believe.  But there are costs to committing the cosmological fallacy.  The most grave 
is that no answer is possible to the three questions.  Sufficient reason is impossible.  In the words of some commentators, our universe seems ``preposterous",  which is another way of saying that its key features are inexplicable within the governing paradigm of explanation.
 
\subsection{Temporal naturalism and law}

By contrast,  a temporal naturalist takes a very different view of the nature of law on a cosmological scale.

\begin{itemize}

\item{} The Newtonian paradigm applies only to subsystems of the universe idealized as isolated systems.  Any use of the Newtonian paradigm is approximate and is limited by the degree to which interactions between the system and its complement may be neglected.  The Newtonian paradigm cannot be applied to the universe as a whole.

\item{}  Any true fact is a truth about the present or the past.   Any verification of a statements truth value must be based on present evidence,  

\item{}  The distinction between laws and states breaks down.   Hence laws may evolve.

\item{}  The future can be at least partly  open. 

\item{}  The evolution of laws allows hypotheses with testable consequences to be put forward to answer the first two questions.  Two examples are given below.  So there is hope for sufficient reason. 

\item{}  Laws may be time asymmetric and irreversible so the third question is 
accessible as well\footnote{For more about this, please see\cite{ECS1}.}.

\end{itemize}

\section{Relationalism and its limits:  relational versus intrinsic properties}

Relationalism is not just a philosophical position, it is a methodological imperative: {\it Progress in physics can often be made by identifying non-dynamical background structures in the description of a subsystem of the universe and replacing it with a real dynamical physical interaction with degrees of freedom outside of that subsystem. }

The paradigmatic example of this is Einstein's use of what he called ``Mach's Principle", by means of which the role of absolute space in defining the distinction between inertial and accelerated motion in Newton's physics is replaced by the action of the distant stars and galaxies, acting through the dynamical gravitational field of general relativity to influence the selection of local inertial frames.  A key step in Einstein's discovery of general relativity was his ``hole argument" which pointed to the role of active diffeomorphisms in wiping out the background structure of the differential manifold, rendering bare manifold points physically meaningless.  The result is that spacetime is NOT identified with metric and other fields on a manifold. It is identified with equivalence classes of such fields under diffeomorphisms.  As already discussed above, in general relativity physical events are defined relationally and contingently in terms of physical effects perceptible there.   

The use of diffeomorphisms in general relativity serves as an example of a general method for eliminating background structure which is to introduce a first, kinematical, level of description which  is encumbered by background structure and then to reduce the description to those engendered by a inimical system of relations by defining physical observables to be those invariant under the action of some group acting on that background structure. This is called gauging away the background structure. 
  Three major examples of this are local gauge transformations in Maxwell and Yang- Mills theory, space-time diffeomorphisms in general relativity and reparameterizations of the string world sheet in string theory.

This may seem like an awkward and indirect way to proceed, but it allows us to write simple, local equations of motion for fields; this would be much more difficult once the naive local fields are eliminated because they are not gauge invariant. Indeed, in all three cases all the physically meaningful observables are non-local.

\subsection{Two relational paths to general relativity: Einstein and shape dynamics}

Since important work of Stachel\cite{Stachel}, Barbour\cite{Barbour} and others the interpretation of general relativity as a relational theory, and the key role of space-time diffeomorphisms in wiping out background structure are well known and appreciated.  But recently there is a new development in the interpretation of general relativity which is highly relevant for the nature of time, which is {\it shape dynamics\cite{SD}}.  This gives a different way
 of defining general relativity by gauging away background structure.   

In the old way developed by Einstein,  space are treated on an equal footing.  Spatial temporal coordinates provide a background structure which is gauged away by imposing spacetime diffeomorphism invariance.  Indeed, one of the things that is gauged away in this story is any distinction between space and time, because there are space-time diffeomorphisms that will turn any slicing of spacetime into a sequence of spaces into any other.   This means there is no meaning to simultaneity.

Barbour has emphasized for years that there is a nagging flaw in the beauty  of this story.  This resides in the fact that there is a big piece of background structure that is preserved in general relativity which is an absolute scale for the size of objects.  We must assume the existence of fixed scales of distance and time which can be compared with each other across the universe.  In general relativity two clocks traveling different paths through space-time will not stay synchronized.  But their sizes will be preserved, so it makes absolute sense to say whether two objects far from each other in space-time are the same size or not.  

You can gauge away this background structure on top of the space-time difeomorphsim invariance of general relativity, but the result will not be 
general relativity\footnote{I know of two ways  to combine space-time diffeomorphism invariance with local scale invariance, one leads to a theory full of instabilities, the other was invented by Dirac and requires extra physical fields to implement\cite{conformalgravity}.}.  The reason  is that imposing another gauge invariance changes the number of physical degrees of freedom.   But the amazing thing is you can get to general relativity by trading the relativity of time of that theory for a relativity of spatial scale, so that the number of gauge transformations, and hence the counting of physical degrees of freedom, are unchanged.   The resulting theory is called shape dynamics\cite{SD}.  

Shape dynamics lacks the freedom to change the slicing of space-time into space and time.  Consequently there is a preferred slicing, i.e. a preferred choice of time coordinate that has physical meaning.  This means that there is now a physical meaning to the simultaneity of distant events.  But physics on these fixed slices is invariant under local changes of distance scale.   

Shape dynamics is not actually a new theory-it is for the most part just a reformulation of general relativity.  Its preferred slices are expressible in the language of general relativity and, indeed, were already known to specialists of classical general relativity.  They are called {\it constant mean curvature} or CMC slices because certain components of curvature are constant on each slice.   The technical statement is that shape dynamics is equivalent to general relativity so long as the space-time has such slices-and most of them do.  (This is modulo space times with black hole horizons, the interiors of which may be different in shape dynamics than in general relativity.)

You might object that these preferred slicings represent a return to a Newtonian conception of absolute time.  But they do not, because the CMC condition is a dynamical condition so that which slices satisfy it depend on the distribution of matter, energy and curvature throughout the universe.  Moreover, 
because the predictions of shape dynamics matches those of general relativity, these preferred slices cannot be detected by any local measurements.  
The slices nonetheless play a role, which can be seen in how the Einstein equations enjoy an impressive simplification when expressed in terms of them.  

Shape dynamics, just by its existence, has two important implications for the present argument. First, the impressive empirical success of general relativity cannot be taken as evidence for claims that the universe is fundamentally timeless, or even that there is no preferred simultaneity of distant events.  These common claims are nullified by the fact that there is an alternative formulation of general relativity that does feature a preferred simultaneity.

Second, the preferred slices of shape dynamics give us a candidate for a global notion of time needed to provide an objective distinction between past, present and future-and hence makes temporal naturalism a possible position to hold-consistent with current scientific knowledge.  

\subsection{Relational purism}

A relational purist believes that once background structures are eliminated physics will be reduced to a description of nature purely in terms of relationships. An important example is the causal set program\cite{CS}, which aims to develop a complete theory of quantum 
gravity-and hence nature-on the basis of an ontology of discrete events, the only 
attributes of which are bare causal relations. These are bare in the sense that 
{\it event A is a cause of event B }
is a primitive. The causal set program denies there are any further properties, P of A and Q of B, such that  {\it P of A causes Q of B. }

The aspiration of the causal set program is to construct the geometry of a lorentzian spacetime approximately satisfying the Einstein equations as emergent only from a discrete set of events and their bare causal relations. To date this has not been 
realized except in trivial cases where the causal set is constructed by randomly sprinkling Minkowski spacetime with discrete events.   

\subsection{Impure relationalism: a role for intrinsic properties}

Completion of the program of eliminating background structures does not imply that there can be no further properties of events except for their causal and other 
relations with other events. In an events ontology, you may eliminate all background structures-as the causal set program very nearly does-and still be left with an event having properties which are not specified when you know all the relations with other 
events. We can call such properties, intrinsic properties. 

Intrinsic properties can be dynamical, in that they play a role in the laws of motion. For 
example, in an events ontology, energy and momentum can be intrinsic properties of events. They can play a role in dynamics and be transferred by causal links. 

This view is realized in the energetic causal set framework\cite{ECS1,ECS2} according to which momentum and energy are fundamental and intrinsic and defined prior to space-time.  Indeed in this approach dynamics is formulated strictly in terms of momentum and energy and causal relations.  Position in spacetime is emergent and comes in at first just as Lagrange multipliers to enforce conservation of energy and momentum at events.

\subsection{Dynamical pairings and relational versus intrinsic properties}

I would like to argue that it is natural to suppose that energy and momentum are intrinsic in a world in which 
space and time are relational\footnote{
Before going on it will be helpful to clear up two terminological confusions.  

{\it Intrinsic versus internal:} If a property of an event is intrinsic it can be defined without regard to any relations to other events. That does not mean it plays no role in the dynamical equations of the theory. Let us reserve the term internal for a property of an event or a particle that plays no role in the laws of physics. Momenta can be intrinsic, but 
it is not internal. Qualia are intrinsic and appear to be internal. 

{\it Structural versus relational:}  By structural properties philosophers seem to mean the same thing that we physicists mean by relational properties. I prefer the term relational as structure seems to denote something static and hence timeless, a structural property 
seems to be one that transcends time or history, but temporal naturalism asserts there may be no such transcendent properties of nature.  Structuralism seems to be a form of timeless naturalism which asserts that what is really real are structures which transcend particularity of  time and place.}.   
This is based on the fact that physics has a particular structure in which spacetime variables are paired with the dynamical variables, momentum and energy: 
This dual pairing is expressed by the Poisson brackets, 
\f
\{x^a ,p_b  \}= \delta^a_b   
\ff
and has its most profound implication in Noether's theorem, which says that if there is a symmetry of a physical system under translation in a physical spatial coordinate $x^a$ then the corresponding momentum $p_a$ is conserved.  Moreover, in the canonical formalism $p_a$ is there generator of translations in $x^a$
\f
\delta x^a = \alpha^b \{ x^a ,  p_b \} = \alpha^a 
\ff
This can be interpreted to say that if position is absolute and so has symmetries (i.e. nature is perfectly unchanged under translations in $x^a$), 
then the corresponding momentum $p_a$ can be defined relationally, in terms of translations in $x^a$.  But note that if space is defined relationally then there can be no perfect symmetry under translations in a space coordinate.  The reason  is that the identity of the indiscernible rules out symmetries because a symmetry is by definition a transformation from one physical state of a system to a distinct state which has identical physical properties.  But Leibniz's principle asserts that no system can have two distinct identical states\footnote{This reasoning does not rule out gauge symmetries which relate different mathematical descriptions of the same physical state.}.  So if space is relational, we loose the relational definition of momentum as the generator of translations.   So if space is relational, momentum can be intrinsic.

We can also turn this around and take the view that momentum is the primary quantity and is intrinsic, and define position relationally  as the generator of translations in momentum space.  This is the point of view taken by the framework of relative locality\cite{RL}. 

Adding intrinsic momentum and energy variables to the causal set description has an immediate advantage which is to resolve a long standing problem with the purist causal set approach, which is to get a low dimensional space-time to emerge from a network of pure causal relations\cite{ECS1,ECS2}.

So I would like to propose that generally we take momentum and energy as intrinsic quantities, defined at a level prior to the introduction or emergence of space-time.    Support for this comes from the Einstein equations: 
\f
R_{ab} - \frac{1}{2}  g_{ab} R = 8¹G T_{ab} 
\ff
The left hand side is composed of geometric quantities that in general relativity are defined relationally.  The right-hand side contains the energy-momentum tensor, which describes the distribution of energy and momentum on space-time.  Ever since Einstein began working on unified field theories generations of theorists, down to late 20th century string theorists, have speculated that progress is to be achieved by reducing the right hand side to geometry as well, so that physics can be expressed in a purely geometric structure.  But perhaps that is mistaken- which would account for its not having worked definitively.  Instead, we can posit that it is the left side that is emergent from a more fundamental description in which energy and momenta are among the primary quantities, perhaps along with causal relations.  

\subsection{The uniqueness of fundamental events}

For a naturalist, the universe, being the totality of all that exists, must be unique. For a relationalist this unique universe must contain all of its causes.    But a 
little known consequence of Leibniz's principle of the identity of the indiscernible is that every elementary event must be unique, in the sense of being distinguishable from every other event in the history of the universe by its location in the network of relations\footnote{The ideas of this subsection were developed in \cite{ECS1,ECS2}.}.

It is most straightforward to make this argument within an ontology in which the history of the universe consists of discrete events whose relational properties are fundamentally causal relations, but the argument can be made in  other ontologies as well.  These events may, as we have just discussed, also have intrinsic properties.  But they have no intrinsic labels, if we want to refer to aparticular event, we cannot just give it a name, such as ``event A", we must give a unique description in terms of its relation to other events, that uniquely distinguishes it from all the others.  And in that description you cannot just say, event A is the one whose immediate causal past is events B and C-for B and C  must be specified relationally as well.  So the description must be based on the past causal set, going far enough into the past that event A's  causal past is distinct from the causal past of any other.  

So the point is that elementary events are not simple to name, because a complete specification in terms of relational properties must contain enough information to single them out from the vast number of other events in the history of the universe.

Given this, let's consider what goes into spelling out the most elementary causal relation, i.e. events B and C are the direct causal progenitors of event A.   Developed in enough detail to relationally describe or address the three events, this is a highly complex statement that carries a vast amount of information.   So elementary events are not simple and neither are elementary causal relations.

It follows that at the fundamental level nature cannot be governed precisely by laws that are both general and simple.  A complete law of nature would have to explain why event A occurred, which means why the events B and C gave rise to a new event.  To give a definite answer to this question would be to pick out which pairs (or small sets) of events give rise together to new events.  As there are vastly more pairs  or subsets that do not have common immediate future events, the question is what makes B and C different from the vast numbers of pairs that do not give rise to new events.  Any explanation for this must be based on what makes those common progenitors different from the other pairs, i.e. it must involve enough information to pick out by means of their relational properties what makes those that are progenitors distinct from those that are not.   

So any complete explanation for the elementary causal process ``B and C together cause A" must make use of information that distinguishes B and C from all the others.  So it cannot be a general and simple law of the kind we usually envision governs elementary events.  Candidates for such laws are discussed in \cite{ECS1}.   

If we are willing to have coarse grained, rather than exact, laws, then we can speak of classes of events.  For example, A is in the class of events that has two immediate progenitors.  These classes require much less information to describe than the individual events, hence we can have simple laws applying to large classes of events.  These may be quite general,  applying, as they do, to large classes of events.  So we see that general and simple laws of the usual kind can emerge from a more detailed description by coarse graining, so that they apply to large classes of events.  

Indeed, nowhere in physics do we have a theory that explains why individual events occur.  Most theories with an events ontology are quantum theories, such as Feynman diagram approaches to quantum field theory, according to which every causal history, with every possible sets of events and causal relations has an amplitude it contributes to the sum over histories. There are, to my knowledge, no deterministic causal set models, instead those causal set models that have been studied take a stochastic or quantum approach to dynamics.  They therefor do not attempt to answer the question of why particular causal sets may occur.  The only exception I am aware of is the model we developed with Marina Cortes\cite{ECS1}.

\subsection{The Newtonian paradigm from the viewpoint of temporal naturalism}

We can summarize the last few points by describing the proper role of the Newtonian paradigm for a temporal naturalist.  

On cosmological scales the universe is unique and laws evolve: so the Newtonian paradigm breaks down. 
On fundamental scales events are also unique  so the Newtonian paradigm breaks down here also.   Events are distinguished by their relational properties and thus must be fundamentally unique: there can be no simple and general laws on the fundamental scale. 

Repeatable laws only arise on intermediate scales by coarse graining which forgets information that makes events unique and allows them to be modeled as simple classes which come in vast numbers of instances.  Hence the Newtonian paradigm works only on intermediate scales. 
 
We can also see from this that intermediate scale physics must be statistical, because similarity arises  from neglect of information.  
It is interesting to wonder whether this might be the origin of quantum uncertainty.  That is, the hidden variables needed to complete quantum theory, if we are to explain why individual events take place, must be relational.  They must arise in adding the information needed to distinguish each event uniquely from all the others.  Note that because the question of distinguishing individual events from others requires a comparison with others, such 
relational hidden variables must be non-local.  

Finally, it may happen that uniqueness might sometimes not wash out on intermediate scales,
leading to a breakdown of lawfulness, arising from novel states or events.  This idea is developed below as the Principle of precedence\cite{precedence}.  

Thus we are left with a compelling and new underseanding of the nature of laws.  Causality is fundamental and prior to law.
The fundamental laws that guide the process of  causation-if they exist at all-cannot be both local and simple.  All laws that are simple and local must arise from a process of coarse graining-hence they are effective and statistical.  Given the distinction between laws that govern and laws that describe, they are closer to the latter. Hence, simple and local laws when they arise, need not be permanent-so there is no barrier to investigating the possibility that they evolve.  
 
\section{The evolution of laws}

The main advantage of temporal naturalism is that it allows the laws of nature to evolve and, as Charles Sanders Peirce understood already
in 1893, this makes possible explanations for why these laws in terms of hypotheses that are falsifiable by doable experiments.   As Peirce put it

\begin{quotation}

{\it `` 
To suppose universal laws of nature capable of being apprehended by the mind and yet having no reason for their special forms, but standing inexplicable and irrational, is hardly a justifiable position. Uniformities are precisely the sort of facts that need to be accounted for. . . . Law is par excellence the thing that wants a reason.
Now the only possible way of accounting for the laws of nature and for uniformity in general is to suppose them results of evolution\cite{CSP-1}."}

\end{quotation}

The important point is that hypotheses about mechanisms that may have operated in the past to evolve the laws can have consequences for observable phenomena, because they are about the past.  This point  is analyzed in some detail in \cite{TR,SU}, so I will only mention below two specific hypotheses which illustrate the point, the principle of precedence and cosmological natural selection.

It is, however, worth mentioning that the idea that the laws of nature have evolved has been voiced by some of the great physicists of the 20th Century.
Paul Dirac, put forward his large number hypothesis according to which the constants of nature evolve and later wondered about a more general evolution of laws.

\begin{quotation}

{\it 
''At the beginning of time the laws of Nature were probably very 
different from what they are now. Thus, we should consider the 
laws of Nature as continually changing with the epoch, instead of as holding uniformly throughout space-time. 
-Paul Dirac"}

\end{quotation}

Even Richard Feynman wondered in an interview whether physics might turn out to be an historical science:

\begin{quotation}

{\it ``
The only field which has not admitted any evolutionary question is 
physics. Here are the laws, we say,...but how did they get that way,
in time?...So, it might turn out that they are not the same [laws] allthe time and that there is a historical, evolutionary, question\cite{Feynmanquote}. 
-Richard Feynman 
         "}

\end{quotation}

Due to experiments partly motivated by the desire to check predictions of Dirac's large number hypothesis, we now have good limits on the evolution of a few parameters of the standard model including $G$ and the fine structure constant, $\alpha$.  These  render it unlikely that the laws have changed much, if at all,  in the present era.  To explain the present laws through a process
of evolution it therefor might help if the initial cosmological singularity is replaced by a bounce from a previous era.  This is an old speculation, going back to Tolman in the 1930s, but there is recently good theoretical evidence that this is a generic consequence of unifying quantum theory with general relativity.  A bounce instead of a cosmological singularity is a universal feature of models of quantum cosmology constructed using the methods of loop quantum gravity\cite{LQC}.  

Nor is it new that such cosmological bounces might be moments when the laws of nature change.  Feynman's Ph.D supervisor, John Archibald Wheeler, speculated that at such events the parameters of the laws of nature might be randomly reshuffled.  He called this ``reprocessing the universe."

The defining idea of temporal naturalism is then not presentism, it is that the laws of nature are not timeless, but the result of processes of change.  This idea has big consequences which we are only beginning to explore.  
My collaborator on this project, Roberto Mangabeira Unger put a provocative challenge before us when he wrote:

\begin{quotation}

{\it `` You can trace properties of the present universe back to the properties it must 
have had at the beginning. But you cannot show that these are the only 
properties that the universe might have had.  . . . Earlier or later universe might 
have had entirely different laws. . . . To state the laws of nature is not to 
describe or explain all possible histories of all possible universes. Only a 
relative distinction exists between law like explanation and narration of a 
one time historical sequence. 
If you are asked what you mean by the necessity of the laws of nature (that is 
to say by the necessity of the most necessary relations), you can legitimately 
respond only by laying out the substance of your cosmological and other 
scientific ideas. People who appeal to fixed conceptions of necessity, 
contingency and possibility are simply confused\cite{Roberto-quote}."}

\end{quotation}

\subsection{The Principle of Precedence}

I would now like to briefly describe two hypotheses about how laws of nature may change.  More about them can be found in the cited references, the key point I want to make is that these hypotheses are testable.  This is evidence for my larger claim that by making laws evolvable in ways that are testable, temporal naturalism makes cosmology more scientific than timeless naturalism, which by keeping laws as timeless and immutable, also puts them beyond explanation by means of testable hypotheses.

The first of these ideas is called the Principle of Precedence\cite {precedence}.  It is an idea about how quantum dynamics might be the result of evolution.

Let me begin with an assertion we know is true: {\it If we prepare and measure a quantum system we have studied many 
times in the past, the response will be as if the outcome were randomly 
chosen from the ensemble of past instances of that preparation and 
measurement. }

Why is this?  Usually we think that that is because a timeless law will act in the future 
as it has in the past. 

But this is a wild idea.  
What kind of thing is a law that lives outside of time but can act in time on 
every material process?  This violates a central tenet of relationalism which is {\it Einstein's principle of reciprocity, }  according to which
if an entity, A, acts on an entity B to alter it, then B must be able to act back on A.  
Besides, how does an electron know it is supposed to follow the electron law rather 
than the quark law? 
There is a radical metaphysical idea at work, making the crazy seem 
obvious. 

There is a less radical assumption: What was just stated is the only law 
of nature needed. 

\begin{quotation}

{\it If we prepare and measure a quantum system we have studied many 
times in the past, the response will be as if the outcome were randomly 
chosen from the ensemble of past instances of that preparation and 
measurement. }

\end{quotation}

We can cross out a few words and simplify this a bit:

\begin{quotation}

{\it   If we prepare and measure a quantum system we have studied many 
times in the past, the response will be randomly chosen from the 
ensemble of past instances of that preparation and measurement. }

\end{quotation}

My claim is that this principle may be the only dynamical law of nature.

Note how parsimonious this idea is.  It gets rid of the metaphysical baggage carried by the conventional view in which there are
laws of nature which do not themselves evolve which however determine the evolution of everything real.  
The ensemble of past events was real, and properties of it can have present truth values.  Hence,  nothing unreal or un-influenceable is 
reaching into the universe to act on the real. 

For now the key point is that this hypotheses is testable because it implies that the evolution of novel quantum states should depart from the usual predictions of quantum 
dynamics-when those involve only the interactions of the system's components.  Our colleagues in quantum information laboratories are learning to work with artificially generated entangled states of many degrees
of freedom.  These have a combinatorial complexity that makes it unlikely that they have ever been produced naturally in the past, so they have no precedents.

I have stated the principle of precedence in an operational language.  Clearly the principle  calls out for being embedded in a deeper theory.
To explain it we need a fundamental theory in which the present and 
past are meaningful.  The search for such a framework is the subject of current work.

But going on,  I would like to mention that once again Peirce was here first:

\begin{quotation}

{\it
`` [A]ll things have a tendency to take habits. For atoms and 
their parts, molecules and groups of molecules, and in short every conceivable real object, there is a greater probability of acting as on a former like occasion than otherwise. This tendency itself constitutes a regularity, and is continually on 
the increase. In looking back into the past we are looking 
toward periods when it was a less and less decided tendency\cite{CSP-2}. "

Charles Sanders Peirce, ÒA Guess at the Riddle,Ó }

\end{quotation}
 
\subsection{Cosmological natural selection}

This hypothesis was described in detail in a 1992 paper\cite{evolve} and my first book\cite{LOTC} as well as in \cite{TR,SU} and review papers\cite{CNS-review},  so I will
be very brief here.  The idea is to recognize that there is a landscape of physical theories analogies to the landscape of genotypes in population biology and mimic in a cosmological scenario the formal description of a population evolving on a  fitness landscape.  The key hypotheses begin by modifying Wheeler's reprocessing scenario

\begin{itemize}

\item{}  Universes reproduce by black hole singularities ``bouncing", giving rise 
to new regions of spacetime.

\item{} At each bounce the parameters of the laws of physics mutate slightly. 

\end{itemize}
Consequently: 

\begin{itemize}

\item{} A typical universe is most likely to come from a parent that had many 
progeny than few. 

\item{} If our universe is typical then it is likely that the laws that govern it haven been tuned to 
increase the number of black holes over the numbers produced by typical laws.  This implies that the parameters of the standard model are near values that extermize the production of black holes locally in the parameter space. 

\end{itemize}

This explains many of the fine tunings of parameters of the standard model\cite{evolve,LOTC}.  This is because, to maximize a universe's production of black holes

\begin{itemize}

\item{} Star formation requires plentiful carbon and oxygen.  This explains the fine tunings of the electromagnetic and strong coupling constants as well as the proton, pion and electron masses in order to stabilize these elements,  

\item{} Supernovas require tuning of weak interactions.  This explains the tuning of the weak interaction scale.

\item{} Gravity must be very weak.  This explains the large ratio of the Planck to the proton masses, which is part of the hierarchy problem.

\end{itemize}
In addition it makes predictions for real observations that have been done in the years since.  Two predictions made in 1992\cite{evolve} are still standing, as is described in \cite{CNS-review}.  

\begin{enumerate}

\item{} The heaviest stable neutron star must be less than twice the sun's 
mass.

\item{} Inflation, if true, must be single field, single parameter.

\end{enumerate}

\section{Time and qualia}

Now I would like to turn to a new subject, which is the implications of our conception of time for the philosophy of mind.  Strawson\cite{GS} 
and Nagel\cite{TN} write of the need for  naturalism to accommodate qualia, or conscious experience, as a natural part of the physical world.  Here I would like to argue that this is much easier to do in temporal naturalism than in  timeless naturalism.

I can begin with two basic observations.  First,  every instance of a qualia occurs at a unique moment 
of time. Being conscious means being conscious of a moment. Being ordered and ``drenched" in time is a fundamental attribute of conscious experience. 

Second, facts about qualia being experienced now are not contingent.  There are no facts of the form, ``If there is a chicken in the road then I am  now experiencing a brilliant red."  

It follows that qualia cannot be real properties of a timelessly natural world,
because all references to now in such a world are contingent and relational.  Nor can qualia be real properties of a pluralistic simultaneity of moments because what distinguishes those moments from each other are relational and contingent facts.  

Qualia can only be real properties of a world where ``now" is has an intrinsic meaning so that statements about now are true non-relationally and without contingency.  These are the case only in a temporal natural world. 

It has been objected that eternalists can see the history of the universe having ``temporal parts" with intrinsic qualities.  This misses the key point which is that any reference to one of those timeless parts in a block universe framework must be contingent and relational, whereas our knowledge of qualia are unqualified by either contingency or relation to any other fact.

That was the short version of the argument.  Here is a longer version:

We have direct experience of the world in the present moment. Just as the fact that we 
experience is an undeniable feature of the natural world, it is also an undeniable feature of the natural world that qualia are experienced in moments which are experienced one at a time. This gives a privileged status to each moment of time, associated to each experience: this is the moment that is being experienced now. This means that we have 
direct access to a feature of the presently present moment that does not require relational and contingent addressing to define it. We can define and give truth values to statements  about ÒnowÓ which are not contingent on any further knowledge of the world. 

How can these facts about nature: that each qualia is an aspect of a presently privileged present moment, that does not require contingent relational addressing to define or 
evaluate, be incorporated into our conception of the natural world? 
This fact fits comfortably in a temporal naturalist viewpoint, because in that viewpoint all facts about nature are situated in, or in the past of, presently privileged present moments and no relational and contingent addressing is required to define those that refer to the present. 

This fact cannot fit into a timeless version of naturalism according to which there are no facts situated in presently privileged present moments, except when that can be defined timelessly through relational addressing.   The same is the case for Barbour's moment pluralism.
 
We can draw a stronger conclusion from this. There is no physical observable in a block 
universe interpretation of general relativity that corresponds to my ability to evaluate truth values of statements about ÒnowÓ, without any need for further contingent and relational facts. The block universe cannot represent ÒnowÓ because ÒnowÓ is an intrinsic property and the block universe can only speak of relational properties. Hence the block universe is an incomplete description of the natural world. 

That is, because qualia are undeniably real aspects of the natural world, and because an essential feature of them is their existing only in the present moment, qualia allow the presently present moment to be distinguished intrinsically without regard to relational addressing. Any description of nature that does not allow ÒNowÓ to be intrinsically defined is  an incomplete description of nature because it leaves out some undeniable facts about  nature. Hence the block universe and timeless naturalism are incomplete, and hence they are wrong.

\subsection{Two speculative proposals regarding qualia}

I would like finally to offer two speculative proposals regarding the physical correlates of qualia.

Panpsychism asserts that some physical events have qualia as intrinsic properties, some of which are neural correlates of human consciousness. But it does not need to 
assert that all physical events have qualia. Might there be a physical characteristic which distinguishes those physical events that have qualia?

According to the principle of precedence which I discussed above, there are then two kinds of events or states in nature: those for which there is precedence, which hence follow laws, and  those without precedence, which evoke genuinely novel events. 
My speculative  proposal the correlate of qualia are those events without precedence. 

It is commonplace to observer that habitual actions are unconscious in people. Maybe the same 
thing is true in nature. Maybe brains are systems where a lot of novel events take place?

Here is a second question raised by pan-psychism: If brains have states which are neural correlates of 
consciousness, but consciousness is a general intrinsic 
property of matter, then what physical properties correlate to qualia? Or, to put it differentially, in what way do the physical attributes of correlates of consciousness vary when the qualities of qualia vary? 

Panpsychists argue that the elements of the physical world have structural properties and intrinsic and internal properties. 
By arguing that matter may have internal properties not describable in terms needed to express the laws of physics, panpsychists reserve a place for qualia as 
intrinsic, non-dynamical properties of matter. 
I would propose to cut the pie up differently. I would hold that events have relational and 
intrinsic properties, but relational properties include only causal relations and spacetime intervals which are derivative from them. Under intrinsic properties I would include the dynamical quantities: energy and momenta, together with qualia. 
I would go further and relate energy and qualia. I would point out that the experienced 
qualities of qualia correlate with changes of energy. Colours are a measure of energy, as 
are tones.

\section{Objections to temporal naturalism}

I would like to close by mentioning some objections which have been offered to the arguments summarized here, and briefly indicate the responses I would offer.
 
\subsection{An objective notion of the present is inconsistent with the relativity of simultaneity}

Temporal naturalism assumes there is an object observer independent distinction between 
present, past and future. This violates the relativity of simultaneity, which is 
supported by ultra-precise tests of special relativity\cite{gio-lee}. 

Furthermore there is the Putnam argument that presentism plus the relativity of 
simultaneity implies eternalism, i.e the block universe\cite{Putnam}. 

But, as discussed above, there is a formulation of general relativity, which is empirically equivalent to the older one,  shape dynamics, that has a 
preferred global simultaneity because it trades the relativity of time for a relativity of scale\cite{SD}. Hence temporal naturalism can be consistent with all the experimental tests of special and general relativity.

\subsection{The metalaw dilemma} 

If laws evolve, there either is a metalaw by 
which they evolve or there is not.. Suppose there is a 
metalaw. We must ask why that metalaw, hence there is danger of an infinite  regress. But suppose that there is no metalaw.  Then there are 
features of the world that are not explained, i.e. sufficient reason is not gained.
In either case a rational explanation of the world is stymied. 

This dilemma is important, but we argue in \cite{TR,SU} that it is in the class of dilemmas that are resolvable by the invention of new
forms of explanation.   We offer several avenues to resolve this dilemma.

\begin{enumerate}

\item{} The regress arises in a particular form of law, it can be avoided if the metalaw does not have the form of the Newtonian paradigm.
For example a statistical form of metalaw, as is posited in cosmological natural selection, by the stipulation that changes in the parameters of the standard model resulting from bounces are small and random, can provide sufficient reason for some features of those parameters.  Such an hypothesis may or may not be explicable in a more detailed metalaw, but need not be to be explanatory.

\item{} We can take advantage of the weakening of the distinction between  state and law to unify them, so that the distinction between them is emergent 
and approximate. A model illustrating this is discussed in \cite{unify-model}.

\item{} We can posit a principle of universality of metalaws, analogous to the principle of universality in computation, according to which any two candidate metalaws that satisfied a reasonable set of conditions would be isomorphic in their explanatory outcomes.  Models of this are explored in \cite{unify-model,unify-universal} 

\item{}Laws may emerge and evolve with the phenomena they describe, as in biology\cite{SU}. 

\end{enumerate}

These are avenues to explore which show at minimum that the metalaws dilemma, rather than being a dead end, may be the key to further breakthroughs.  We would suggest that the direction of theoretical cosmology in this century will be determined by which turns out to be the right resolution of this dilemma.

\subsection{Internally inconsistent}

Some philosophers offer arguments that temporal naturalism is inconsistent.  Some of these date back to the old presentism-eternalism debate.  Nonetheless, they are at best inconclusive, because they tend to err by assuming what is to be proven.  Temporal naturalism may be wrong, but it is a logically consistent position to take.  Its validity will be determined by the success of failure of its empirical agenda, i.e. the confirmation of predictions which follow from particular hypotheses about the evolution of laws.  

Here is an example of such an argument and why it fails.

Temporal naturalism asserts that the presently present moment has a special status. But it also admits that 
there were moments in the past and will be other moments in the future. {\it But all these 
other moments will have or have had that same special status with respect to observers 
at those times.} Hence any special status claimed for ÒnowÓ must apply to all moments. 

This argument fails because it does not respect the assumptions of temporal naturalism, within which the italicized 
sentence either has no truth value, if it refers to the future or does not demonstrate that the past is the same as the present, if it refers to the past. With regard to the future it has no truth value because 
no statements about the future have truth values. 

With regard to the past it may be true that past observers gave the same special status to their then present moments that present observers
give to our present moments but this does not make the past the same as the present.  All truth values must be evaluated relative to the
present and we evaluate evidence about past moments differently than we evaluate statements about present moments.
The point is that the special status that temporal naturalism gives to the present can only be claimed or evaluated in that present. It is
not transferable.
There may be evidence in records of past observers but this is not the 
same as the evidence that comes from the direct apprehension of ``now". Hence while 
we believe past moments may have existed, that belief is based on indirect evidence; we 
cannot assign that belief the same status as our direct knowledge of the present. 
To put this in terms developed above, the italicized sentence requires contingent, relational 
addressing to make sense of claims about the past, which is different from our 
knowledge of the present. 
 
Another argument sometimes given against the logical coherence of presentism is that no meaning can be given to the ``flow of time" because
any claim that time ``flows" has to answer the question of what determines the rate of time's flow.  But, so the argument goes, it makes no sense to
talk about a ``rate of flow of moments."   

To me this is a pseudo problem.  Temporal naturalism has no need to speak of the rate of flow of time.  Nor does this question need to be answered. In relativistic physics we have a perfectly cogent notion of proper time, yet no one ever wonders about its rate of flow.  Or rather the answer is either trivial, i.e. the rate of proper time with respect to itself is unity, or arbitrary, as in the rate of flow of proper time with respect to an arbitrary parameterization of a world line.  

A third argument against the cogency of presentism starts with the query that the instant cannot be instantaneous, because all perceptions of moment actually require a duration.  The present must then be thick, i.e. comprising a duration, which means we have to answer the question as to how thick is the present?  

This seems correct and in \cite{ECS1} we propose a precise specification of a thick present within an ontology of events.  
 
\subsection{Can causality be prior to law?}

The standard view ties causality to timeless law: causal implication is explained by being actions of law.  But if laws evolve, while present events nonetheless
 continually give rise to future events, causality must be prior to law.  Can we maintain a notion of causality as prior to law.

A notion of causal inference prior to law is developed in \cite{ECS1}. Causality is envisioned as a consequence of what we call
{\it  the activity of time}:  this is to bring future events into being 
out of present events. Whereas any law is contingent and temporary, this activity of 
time can be posited to be the one aspect of nature that is inherent and immutable. 

Another reason causality must be prior to law stems from the principle of the identity of the indiscernible.  This implies that all events are 
unique, ie its properties, both relational and intrinsic, must be distinguishable from those 
of every other event. Thus the simple statement A and B are causes of C is 
incredibly complex because the full specification of A,B and C require vast amounts 
of information. But if each event and each causal influence are incredibly complex 
then there are no laws governing the microscopic scale that are both simple and 
general. 

We see already in general relativity, the incredible complexity of contingent observables 
needed to specify and describe local events.

\subsection{Can eternalists believe in the evolution of laws?}

The point is sometimes raised that an eternalist can believe that laws evolve.  In a certain sense this is true but this misses the point because it rests on a change in the meaning of the word ``evolve."  What an eternalist or block universe person means by an  evolving law is a dependence of the law on a time parameter which could be described timelessly, i.e an example would be a hamiltonian $H(t)$ depending on $t$, the age of the universe.  But this is just a particular form that a timeless law could take.   

This is also a  confusion brought about by the difference between the physicists' and the philosophers' notions of the block universe.  That is, an eternalist who is happy with the very weak philosophers's block universe can admit evolving laws, but the temporal naturalist, constrained by the stronger physicists' block universe, cannot.

\subsection{What is the status of mathematics? }

Temporal naturalism is incompatible with Platonic and Pythagorean views of 
mathematics and its relation to physics.  Platonism is the view that mathematical objects exist within a separate reality.  No naturalist
can be a Platonist as naturalism asserts that there is only one mode of existing which is that of physical existence.   

Pythagoreanism is the view that the world, or else its history, is really a mathematical object.   It is clear that a temporal naturalist cannot be
a pythagorean.  In fact what we can argue that if one only posits the reality of the experience of the present moment one can deduce that
the world and its history are not isomorphic to any mathematical object, MO: 
For, were it so, every property of nature would have a corresponding property of MO. 
Here is one property of the world that has no correspondent in any property of any 
mathematical object: that it is always some moment which is one of a succession of 
moments. 
Properties of mathematical objects are true timelessly, ie if true they are always true. 

So, for a temporal realist, the effectiveness of mathematics in physics is limited to necessarily incomplete 
modeling of subsystems of the universe, hence the effectiveness is reasonable. Wigner's worry about the
``unreasonable effectiveness of mathematics is physics" is resolved by reducing our expectations for what mathematics can represent from a
correspondence with every property of the whole universe to a very effective, but approximate, description of the observed behaviours of subsystems of the universe.  

\section{Conclusions}

The main theme of this essay is that the concept of time we hold has a great influence on how we understand the core elements of natural philosophy such as state and law.  I have shown that there are different versions of naturalism which differ on the conception of time, state and laws in ways that have strong implications for the future direction of physics and cosmology.  I argued that one of these, timeless naturalism, is at a dead end as a framework for thinking about cosmology.  I gave several reasons for this, one of which is because the why these laws and why these 
initial conditions questions are unanswerable within it. On the other hand, I argued that
temporal naturalism offers avenues to resolve these questions and hence is the way 
to continued progress of science. 

Most of this essay briefly introduced arguments that are developed in much more detail in \cite{SU,TR}.  But I used the opportunity to introduce
some new themes which arose in recent work.  One is a cluster of ideas connecting the uniqueness of elementary events to the irreversibility of fundamental laws and the non-relational, intrinsic character of energy and momenta.  

I also suggest that temporal naturalism may be useful for philosophy of mind, and particularly for those interested in Chalmer's hard problem of consciousness\cite{Chalmers}, which is the relationship between the qualia of experience and the physical world.  I argued that
belief that qualia are real and are a part of nature is closely related to the belief 
that the present moment is real. 
Both are direct apprehensions. 
Both are internal and intrinsic non-relational aspects of nature. 
Hence, temporal naturalism is qualia friendly.

\subsection{What is at stake?}

I close by looking more broadly at what is at stake in the debate between timeless and temporal naturalists\footnote{The choice for science and its implications are the subject of an essay\cite{me-brick}.}.  The two views explain a certain clustering of views scientists and philosophers often, but not uniformly, tend to hold on diverse issues.  Thus, I have observed that those who believe in strong artificial intelligence (AI), tend also to believe that
nature is a computation, in the illusory character of qualia and the passage of time,  in the many worlds interpretation of quantum theory as well as in anthropic multi-verse cosmologies.  Those who tend to  be realists about time and its passage are more likely to believe that qualia are real, that
quantum mechanics is incomplete and to be skeptical both about strong AI and anthropic multiverse cosmologies.  While exceptions exist which show that  these correlations are 
not perfect, their ubiquity suggests there is something to explain.  

I would suggest that what is at root in this divergence of scientific world views is the disagreement we have discussed here between temporal and timeless naturalism.  

Timeless naturalism is associated broadly with the views that: 

\begin{itemize}

\item{} The future is determined. The world is nothing but elementary particles with 
timeless properties moving and interacting according to timeless laws.

\item{} Novelty, 
agency, purpose, intentionality, will are all illusions. 

\item{} The universe is or is equivalent to a computer, classical or quantum.

\item{} Hence strong AI: persons are also perfectly emulate-able as computations.

\item{} Laws and initial conditions are inexplicable except via the anthropic principle in a 
multiverse, ie sufficient reason  is impossible and modern science is at an end. 

\end{itemize}

Temporal naturalism holds instead that: 

\begin{itemize}

\item{} The future is partly open, in that laws evolve on cosmological and perhaps also 
quantum scales.

\item{} The long term evolution of the universe is not computable.  The history of the universe is not isomorphic to any mathematical object.

\item{} Persons have properties that cannot be captured by any computation. 
The brain is a physical system, but not a programmable digital computer.

\item{} Laws and their initial conditions may be explained by falsifiable hypotheses as to 
their mechanisms of evolution. ie there is much science still to do. 

\end{itemize}

These divergent views confront each other within the communities of scientists and underlie debates within fields as diverse as cognitive science and cosmology.  I hope I have said enough in this essay to suggest that temporal naturalism is the naturalism of the future.

\section*{Acknowledgements}

I would like first of all to thank Emily Grosholz for her invitation to contribute this paper to the special Issue of  Studies in History and Philosophy of Modern Physics on ÒTime and CosmologyÓ.  I am most grateful to Roberto Mangabeira Unger for our now seven years of  collaboration on these issues leading to  our joint book\cite{SU}.  I am also grateful to Marina Cortes for our recent collaboration on the work reported above\cite{ECS1,ECS2}. 

This paper has benefited greatly from four opportunities to present versions of it to audiences of philosophers. I want to thank Simon Saunders for an invitation to present an early form of this essay as a talk at the Oxford philosophy of physics group and to Simon, Harvey Brown and other members for astute critical comments.  I owe a similar debt to Hedda Hassel for an unexpected invitation to present the argument on qualia to the workshop PANPSYCHISM, RUSSELLIAN MONISM AND THE NATURE OF THE PHYSICAL held  August 23rd-24th 2013 at the University of Oslo.  I am especially indebted to members of the workshop for astute comments including especially Galen Strawson, David Chalmers,  Hedda Hassel, Bill Seager and Philip Goff.  I am equally indebted to Alisa Bokulich and her colleagues for the organization of a symposium on Time in Cosmology within the Boston Colloquium for the Philosophy of Science, and specially to the other speakers, Jenann Ismael, Chris Smeenk and Roberto Mangabeira Unger, as well as to Peter Bokulich and John Stachel for their critical comments.  And I learned immeasurably from a workshop on Space, Time and Experience, organized by David Chalmers where this paper was read and critiqued by David Albert, L.A. Paul, Tim Maudlin, and other participants.

Conversations over many years with Jaron Lanier have been crucial for shaping my views on the implications of temporal naturalism.  I also am indebted to Julian Barbour, Fotini Markopoulou, Huw Price, Carlo Rovelli and Steve Weinstein for many discussions, arguments and challenges.  
This research was supported in part by Perimeter Institute for Theoretical Physics. Research at Perimeter Institute is supported by the Government of Canada through Industry Canada and by the Province of Ontario through the Ministry of Research and Innovation. This research was also partly supported by grants from NSERC, FQXi and the John Templeton Foundation.


\end{document}